\documentclass{iaus}
\usepackage{graphicx,amssymb}

\newcommand {\m} {$\mu {\rm m}$}

\title[~~EBL constraints from gamma-ray absorption studies] 
{Constraints on the Optical-IR extragalactic background 
from $\gamma$-ray absorption studies}

\author[Luigi Costamante]   
{Luigi Costamante$^{1,2}$
}

\affiliation{$^1$ Stanford University, USA \& INAF-OAB, Milano, Italy \\
$^2$ Now at: Dept of Physics, University of Perugia, Perugia, Italy  \\
email: {\tt luigic2011@gmail.com}}

\pubyear{2011} 
\volume{284}  
\pagerange{1--12}
\jname{The Spectral Energy Distribution of Galaxies}
\editors{R.J. Tuffs \&  C.C.Popescu, eds.}
\begin{document}

\maketitle

\begin{abstract}
Very high energy (VHE, $\gtrsim0.1$ TeV) gamma-rays from extragalactic sources, interacting 
by $\gamma-\gamma$ collisions with diffuse intergalactic radiation fields, 
provide an alternative way to constrain the diffuse background light,
completely independent of direct measurements. 
The limits depend however on our knowledge of the physics of the gamma-ray sources. 
After clarifying the interplay between background light and VHE spectra, 
I summarize the extent and validity of the obtainable limits, and where 
future improvements can be  expected.
\end{abstract}

\firstsection 
              %

\section{Introduction}
VHE $\gamma$-rays\footnote{In the following. 'VHE' and 'TeV' will be  used as interchangeable terms,
meaning about $\pm$1 decade around 1 TeV. The same for 'HE' (high-energy, $>$0.1 GeV) and 'GeV' terms.
In fact, TeV and GeV are the energies at which atmospheric Cherenkov telescopes and the 
Fermi-LAT detector have the best sensitivity and resolution, respectively, for most astronomical spectra.}
from extragalactic sources provide an alternative 
and completely independent way, with respect to direct measurements,
to probe the diffuse Extragalactic Background Light (EBL, from UV to far infrared wavelengths;
see e.g. \cite{hauser}).
This approach is based on the study of ``absorption" features imprinted on the GeV-TeV spectra  by
the interaction of $\gamma$-rays with EBL photons  through the pair-creation process 
($\gamma\gamma\rightarrow e^{+}e^{-}$; see e.g. \cite{felix_icrc} and refs therein).     
Blazars represent a very useful class of $\gamma$-ray beamers, 
being numerous over a wide range of redshifts, very luminous (enhanced by relativistic beaming) 
and long-lasting sources.
However, they are far from being standard candles, and therefore the constraints
on the EBL are {\it always} heavily dependent on our understanding of the blazar 
intrinsic emission and physical properties.

Nevertheless, it is possible to derive meaningful constraints if the intrinsic blazar properties 
implied by certain EBL spectra are very far from the known or expected range,
and/or if they become inconsistent with other aspects of the blazars spectral energy distribution (SED).
Observations in the last 5 years allowed a substantial progress, 
providing the stringest constraints to date.
Here both the main results and the limits of their validity are discussed. 

\begin{figure}[t]
\begin{center}
 \includegraphics[width=7.4cm]{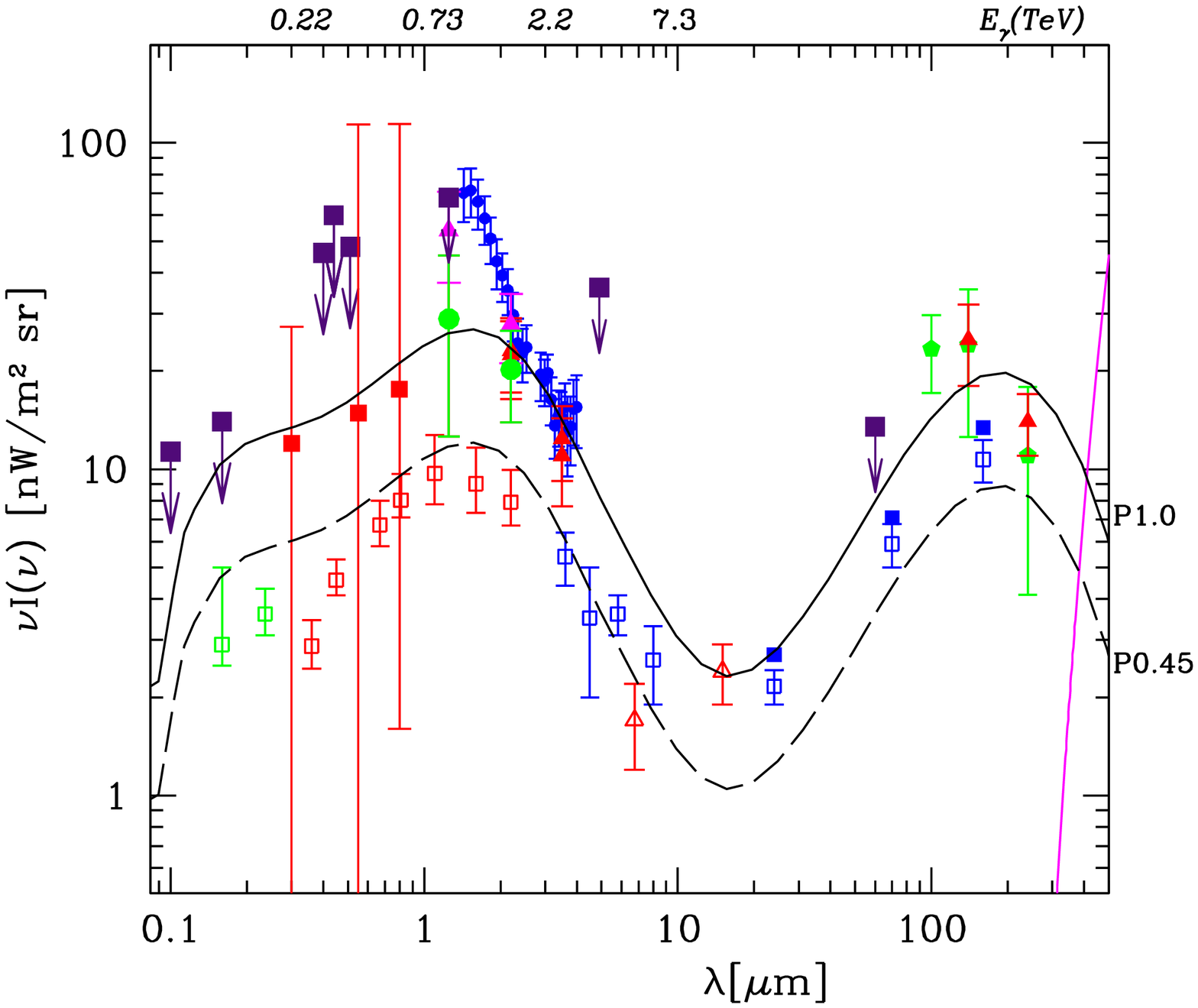} 
 \includegraphics[width=6cm]{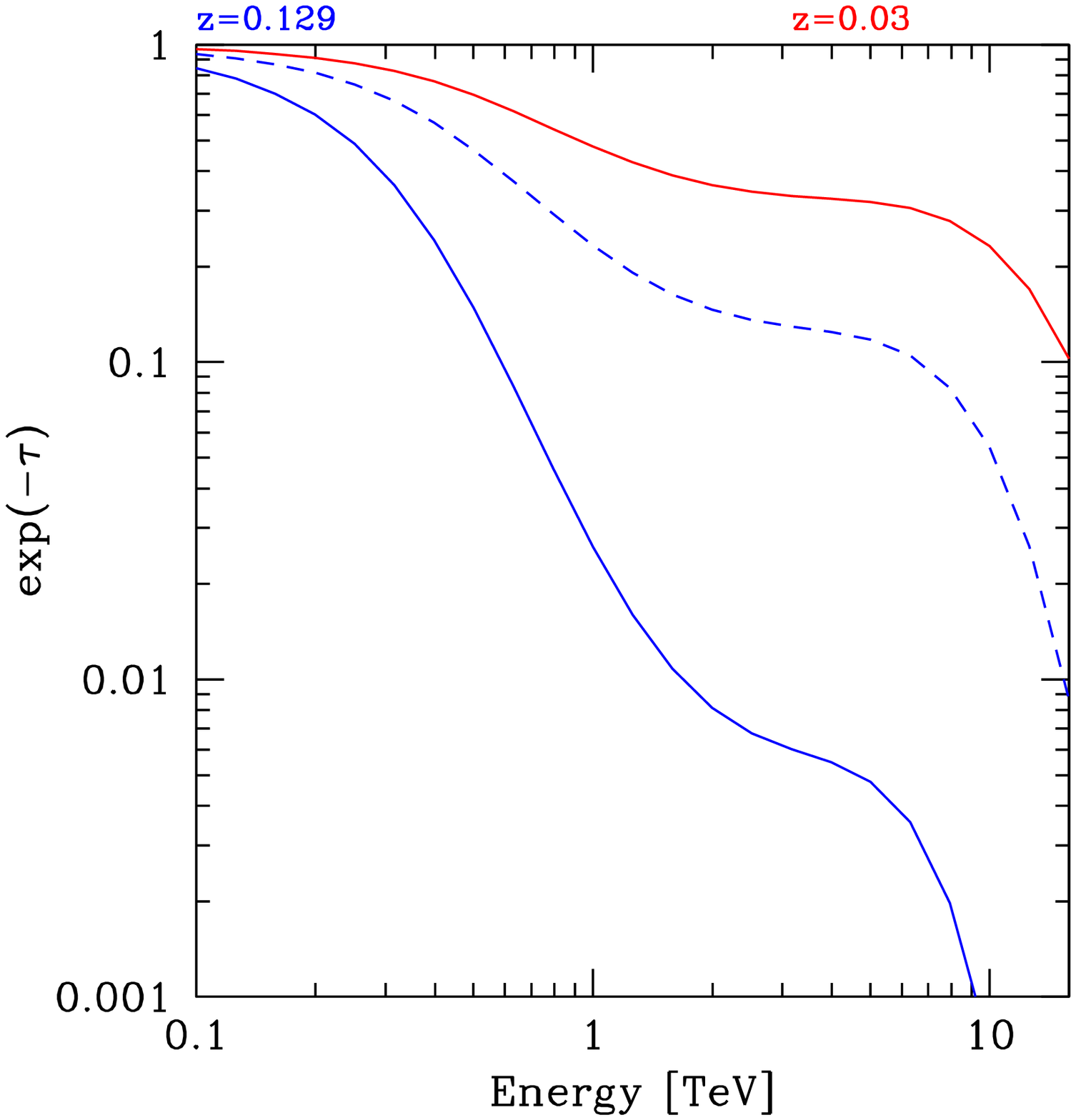} 
\caption{Left: SED of the EBL, as was discussed in year 2005 (from \cite{nature}).
Open symbols show lower limits from galaxy counts, while filled symbols correspond to direct estimates.
On the upper axis it is plotted the TeV energy corresponding to the peak of the $\gamma-\gamma$ cross section.
The two curves (identical except for normalization) are drawn as simple shapes reproducing 
the main EBL features.
Right: the attenuation curves $e^{-\tau}$ corresponding to the EBL curves in the left panel
(lower solid and dotted lines), and at two different redshifts for the upper EBL curve 
(dark and grey solid lines). The attenuation curves represent directly the shape of an observed 
VHE spectrum,  if the initial spectrum were a power-law parallel to the upper axis. 
All calculations performed assuming $H_0=70\;\rm km/s/Mpc$, $\Omega_m=0.3$, $\Omega_{\Lambda}=0.7$. }
   \label{fig1}
\vspace*{-0.6 cm}
\end{center}
\end{figure}

\section{Diagnostics: how absorption deforms TeV spectra}
The convolution of the $\gamma$-$\gamma$ cross section with the EBL spectrum yields
an energy-dependent attenuation. Because the resulting optical depth $\tau$ mostly increases
with $\gamma$-ray energy, the observed $\gamma$-ray spectrum emerging at the end of the travel path
is steeper than the initial, intrinsic spectrum emitted by the source 
(namely, the photon index $\Gamma_{obs}\geq\Gamma_{int}$).
This can be seen in Fig. \ref{fig1}, where the attenuation factor $e^{-\tau}$ is plotted 
as a function of energy.
Limits on the EBL can thus be derived if the intrinsic spectrum required by a particular
EBL intensity (or shape) is too anomalous --typically too hard-- with respect to the known physics 
of the source (e.g. \cite{costamante1}).
Several points can be noted from Fig. \ref{fig1}: 

{\bf 1)} different EBL wavebands (having different slopes) affect each part of the $\gamma$-ray spectrum 
in a different way,  yielding a complex deformation shape. 
However, over some ranges (namely 0.2-1 TeV and 2-8 TeV), it approximates a power-law shape.
That is, if the intrinsic spectrum is a power-law, the observed spectrum can be well fitted 
by a power-law of steeper index. 

{\bf 2)} The amount of steepening $\Delta\Gamma$ increases both with distance and EBL intensity.  
Thus redshift gives leverage: the same uncertainty in EBL flux $\Delta F_{\rm EBL}$
causes a higher $\Delta\Gamma$ at larger distances. More distant sources
provides therefore more sensitivity  for EBL constraints, but at the same time they suffer
stronger attenuation, yielding much lower statistics. 
As a result, there is an optimal range of redshifts that provides the best compromise
between EBL-sensitivity and statistical error,
and thus gives the most stringent EBL limits. With the present detectors, it is around $z\sim 0.15-0.25$.

{\bf 3)} If the EBL number density $n(\epsilon)\propto \epsilon^{-\beta}$ (i.e. the EBL SED 
$\propto \lambda^{\beta-2}$), the optical depth $\tau(E_{\gamma})$ becomes 
$\propto E_{\gamma}^{\beta-1}$. Where $\beta\approx 1$, the optical depth  
becomes therefore independent of energy  (\cite{felix_icrc}). 
In such case {\it there is no steepening}: 
the absorbed spectrum reproduces the original shape, simply attenuated by a constant factor.
This is what partly happens in the EBL waveband 3-10 \m,  according to all most recent
EBL calculations (e.g. \cite{franceschini, dominguez, gilmore11, kneiske}). 
It causes a flattening feature in the attenuation curve between $\sim$1-2 and 8-9 TeV,  
where the observed $\gamma$-ray spectrum partly recovers its intrinsic slope 
and thus can appear harder than in the 0.2-1 TeV band 
\footnote{In fact, $\gamma$-$\gamma$  absorption can make a gamma-ray spectrum even {\it harder}  
than originally emitted, when $\beta<1$   
(e.g. for a planckian distribution). 
This is indeed an effective way to explain very hard $\gamma$-ray spectra 
(\cite{intabs}).}

{\bf 4)} Therefore, {\it NO cutoff is produced by EBL absorption between 0.2 and 8 TeV} ! 
Any cutoff or steepening seen in this energy range should be intrinsic to the source spectrum.
Fig. \ref{fig1}  also clarifies why almost all "data points" reported on the so-called ``gamma-ray horizon" plots
in literature (showing the $E_{\gamma}$ at which $\tau=1$ as a function of redshift) are basically meaningless: 
there is no way to measure an EBL cutoff  between 0.1 and 10 TeV using only VHE data. 
The only two EBL cutoffs are between the HE and VHE bands, i.e. between 20 and 200 GeV 
(thus requiring a Fermi-LAT spectrum),  or above 8-10 TeV (for which only two blazars have been detected 
with sufficient statistics so far).   

{\bf 5)} The $\gamma$-ray steepening is determined by the difference in  optical depth 
between the two ends of an observed VHE band.
Therefore any spectral steepening   (and consequently any EBL upper limit) can always be 
counteracted and canceled  by an appropriate increase of the  optical depth (i.e. of the EBL flux)
at the lower $\gamma$-ray energies, 
so to equalize the optical depths.   In the 0.1-1 TeV range, this is obtained by increasing the UV flux 
with respect to the NIR flux (see \cite{nature}).  
Therefore, {\it any limit on the EBL always depends on the assumption on the UV flux}, by definition.
All  claims of ``new" methods for ``model-independent" or ``shape-independent"
EBL limits (e.g. \cite{mazinraue,hofmann})
are therefore essentially incorrect, even if their (often hidden) assumptions are reasonable and 
in fact commonly used (e.g. assuming that the UV flux is always lower than the Opt-NIR flux).

\section{Optical-NIR constraints}
Until 2005, the large uncertainty especially around 0.8-3 \m\  
(a factor 5-10$\times$ between lower limits and direct estimates) caused a fundamental ambiguity
in the interpretation of blazars gamma-ray spectra.  An observed TeV spectrum 
could be the result either of a hard intrinsic spectrum attenuated by a high-density EBL, or 
of a soft intrinsic spectrum less absorbed by a low-density EBL.  In both cases, the required intrinsic
TeV spectrum was well within the typical range of spectra 
shown so far by blazars, so no useful constraint could be derived (with the exception 
of the 2-10 \m\ range, see \cite{hegra1426}).   
This ambiguity essentially undermined our capability to study blazars, 
since it corresponded to large differences both in energy and luminosity of their SED $\gamma$-ray peak.

\begin{figure*}[t]
\begin{center}
 \includegraphics[width=0.44\textwidth]{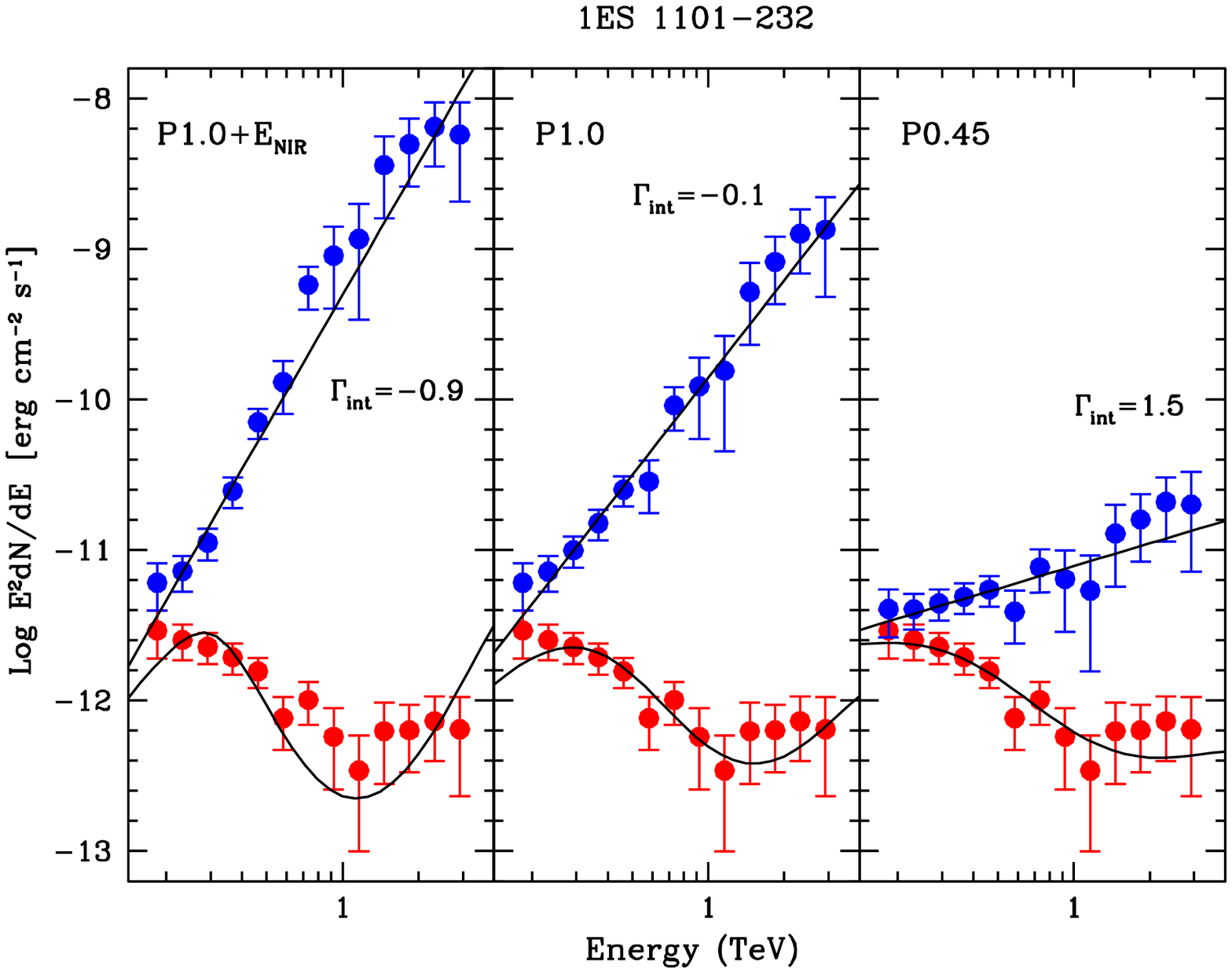} 
 \includegraphics[width=0.44\textwidth]{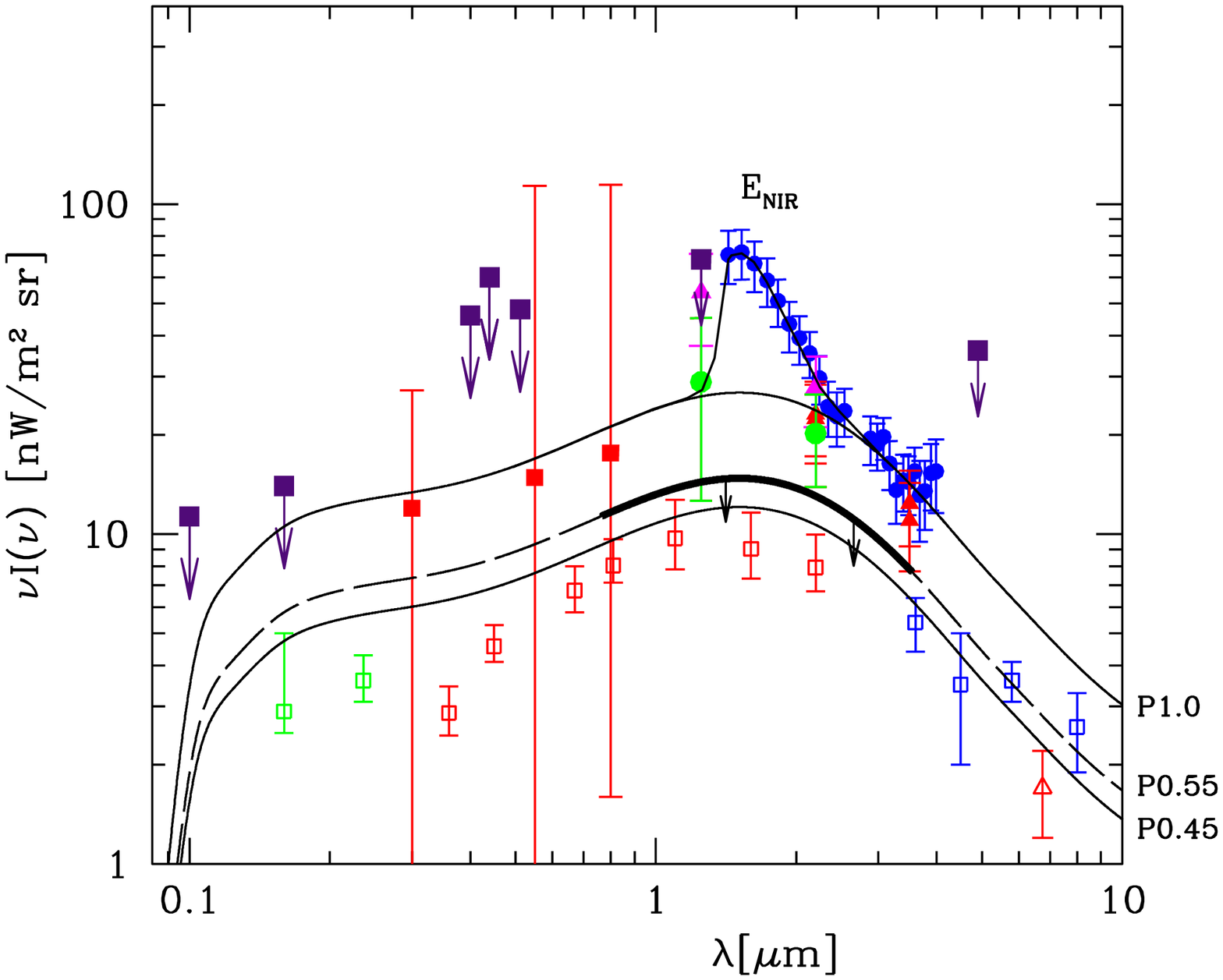} 
 \caption{Left: The HESS spectra of 1ES\,1101-232, corrected for absorption 
with three different EBL SEDs, as labelled (from \cite{nature}).
Lower points (red): observed data. Upper points (blue): absorption-corrected data.
The lines show the best fit power-laws to the reconstructed spectrum, and the corresponding 
shape after absorption.
Right: upper limit for the EBL (black marked region), 
from the assumption of a blazar spectrum $\Gamma\geq1.5$. }
   \label{fig2}
\end{center}
\end{figure*}

In 2005, however, the HESS observations of the BL Lac objects 1ES 1101-232 ($z$=0.186) 
and H 2356-309 ($z$=0.165) provided a fundamental breakthrough (\cite{nature}). 
The observed $\gamma$-ray spectra (detected between 0.2 and 1-3 TeV) 
were much harder than expected for their redshift, 
implying extremely hard intrinsic spectra ($\Gamma_{int}\lesssim 0$) in case of high EBL densities
(Fig. \ref{fig2}).
Such hard spectra were never seen in the closer, less absorbed objects and 
were at odds with all the known blazar physics and phenomenology.  
They were also not supported by same-epoch multiwavelength observations of their SED
(\cite{1101}).
A low EBL intensity could instead accomodate all the new results within the typical range 
of blazar properties. Assuming that the intrinsic spectra were not harder 
than $\Gamma$=1.5 an upper limit could be derived, which resulted very close to the lower limits 
given by the integrated light of galaxies (details in \cite{nature}).
This result is now further corroborated by several other sources and observations
(e.g. \cite{0347, 1218, magic}).
 
This result has three main consequences:  1) it pins down the origin of the EBL,
showing that at these wavelenghts it is strongly dominated by the direct starlight from galaxies, 
excluding a strong contribution from other sources like Pop-III stars (e.g. \cite{santos});
2) it means that the intergalactic space is more transparent to $\gamma$-rays than previously thought,
thus enlarging the $\gamma$-ray horizon; 3) it strongly reduces the ambiguity on blazars TeV spectra.

\subsection{The (in)famous limit of $\Gamma$=1.5.}
There is a lot of confusion in literature on the validity of the $\Gamma$=1.5 hardness limit
and on the reasons of its adoption. 
It is important to understand that  {\it this is NOT the hardest possible theoretical spectrum}
in blazars, and was never introduced as such (see \cite{nature}).
Indeed there are many possible mechanisms to produce extremely hard gamma-ray spectra in blazars, 
such as  bulk-motion comptonization (\cite{felix_icrc}), 
internal absoption on narrow-banded photon fields (\cite{intabs}), 
uncooled particle acceleration spectra or fine-tuned shock acceleration 
(yielding however $\Gamma\sim1.2$, e.g. \cite{steckeracc}), 
a low-energy cutoff in the particle distribution at very high energy (e.g. \cite{katar, lefa}),
or relativistic Maxwellian distributions (e.g. \cite{henri, lefa}). 

The $\Gamma$=1.5 limit is a reference value, a benchmark that so far marks  
{\it the borderline between reality and speculation}. 
Observations do show that blazars can have photon spectra  with $\Gamma\gtrsim1.5$
(directly seen in synchrotron or inverse Compton emissions), and  these can be produced with standard 
shock acceleration and cooling mechanisms without invoking special conditions or fine-tuning.
On the other hand, much harder spectra have never been significantly detected 
in blazars so far, at high particle energies, neither directly or by synchrotron emission.
The alternative scenarios, though not excluded by observations, are not suggested 
by them either, and in some cases require extreme fine-tuning to avoid conflicts with
multiwavelength SED data.

Furthermore, the $\Gamma$=1.5 value should not be considered a sharp, hard limit.
At any given redshift, EBL absorption establishes a one-to-one relation between the observed
and intrinsic $\gamma$-ray spectral slope. Thus a different assumption shifts the EBL limit accordingly,
but only a large change (e.g.  $\Gamma_{int}<<1$) can effectively impact
the conclusion of a low EBL. 
For example, in the case of 1ES 1101-232, a shift of $\Delta\Gamma=0.3$ 
(e.g. by assuming $\Gamma_{int}\sim 1.2$)  affects the EBL limit by no more than 10\% (\cite{nature}).
The Opt-NIR upper limit is therefore robust to such changes in blazar assumptions, 
which are of the same order of the statistical and systematic uncertainty in the $\gamma$-ray spectra.

The only relevant issue left, therefore, is to assess --theoretically and observationally--
if blazars do have emission components
characterized by {\it extremely}  hard spectra (e.g. $\Gamma=0.7$ or less) or not, and 
when they produce them.

\begin{figure*}[t]
\begin{center}
 \includegraphics[width=0.44\textwidth]{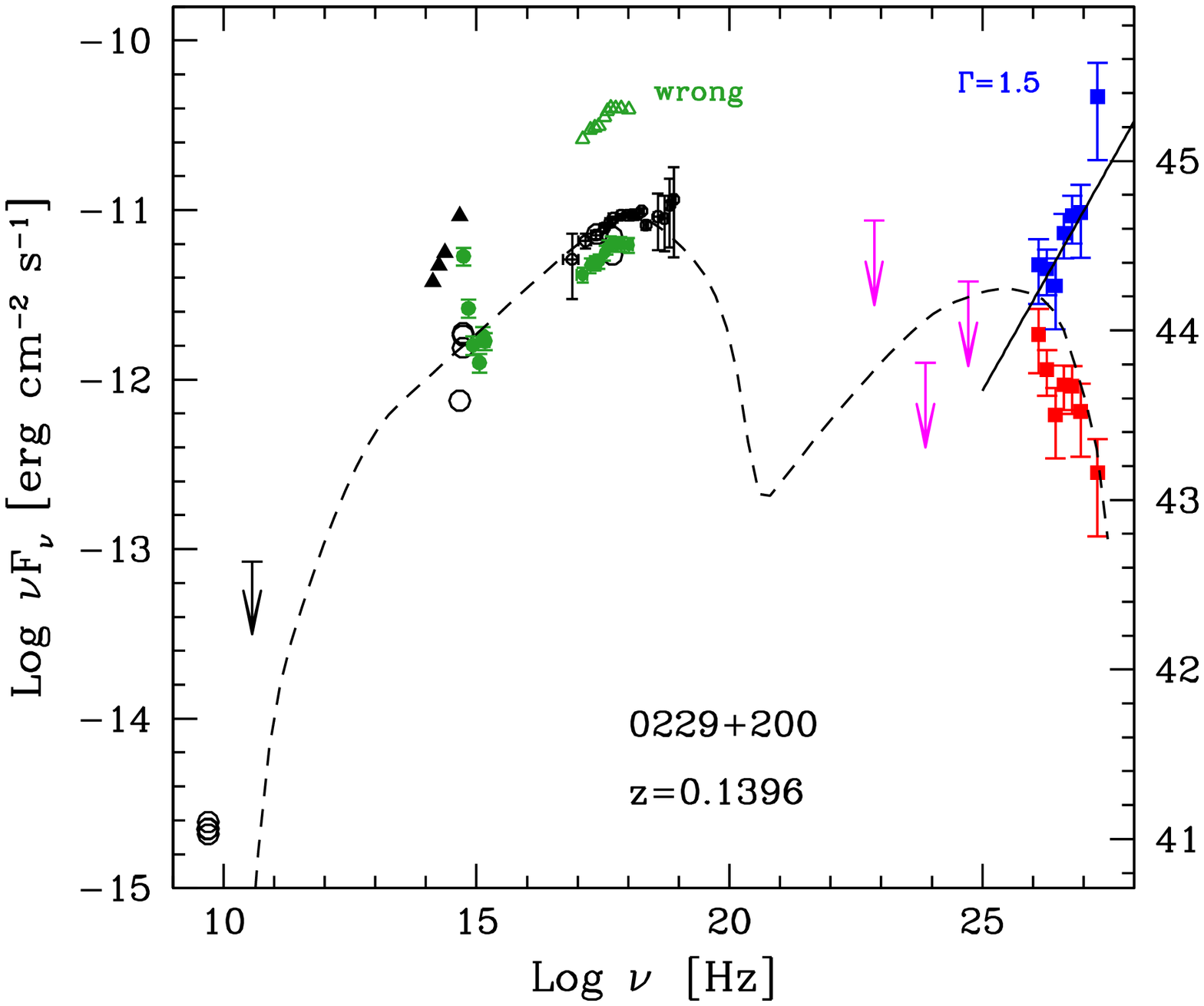} 
 \includegraphics[width=0.44\textwidth]{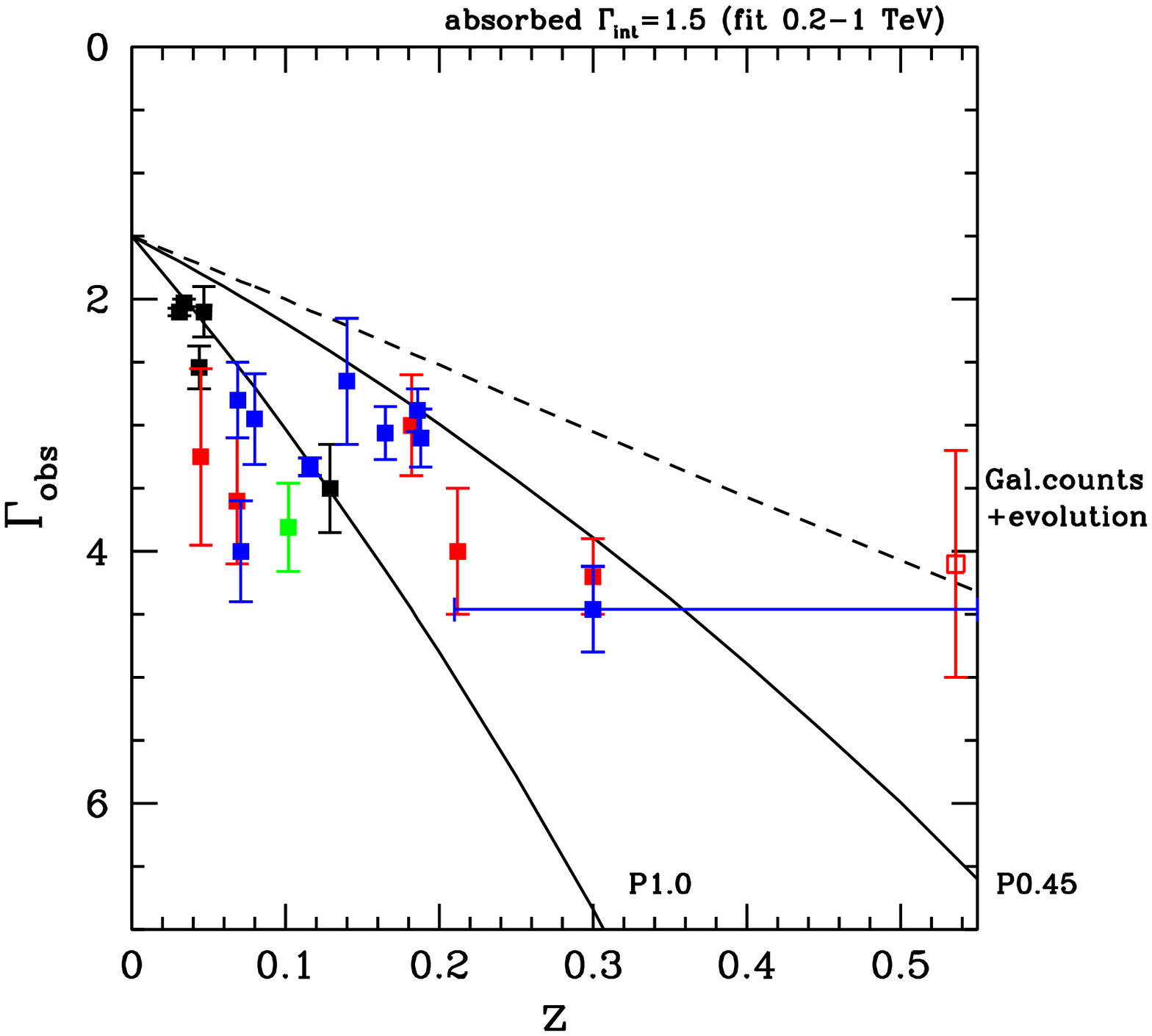} 
 \caption{Left: the SED of 1ES\,0229+200, showing  the correct 
X-ray data from Swift (filled green/grey points), at much lower flux 
than the wrong X-ray data (open triangles) used in \cite{tav0229}.
Right: 
observed VHE photon indices of all detected TeV blazars 
as a function of redshift.  The lines represent the expected observed (i.e. absorbed) photon index
of a source with an intrinsic spectrum $\Gamma=1.5$,
when fitted in the range 0.2-1 TeV, for different EBL levels (as labelled). 
The separation of the data from the lines at each redshift shows how much the intrinsic spectrum is 
harder or softer than 1.5. Adapted from Costamante 2006.} 
   \label{fig3}
\end{center}
\end{figure*}

\section{Why a low EBL density is still the most likely solution.}
Despite the several possible scenarios for producing very hard spectra in blazars,
there are two main reasons why a low EBL density seems still the correct solution.

The first concerns the emitting particles spectrum. Most of these scenarios
invoke very hard particle spectra,  either as a pile-up/Maxwellian distribution or 
as a low-energy cutoff, 
and suppressing radiative cooling. 
These mechanisms look feasable  even in a one-zone synchrotron self-Compton
scenario for blazars (see \cite{lefa} for a recent comprehensive discussion).
However, since the synchrotron emission (which in these objects correspond to 
optical--X-ray energies) traces directly the particle spectrum,
such hard features should appear in the synchrotron spectrum as well, at some energies and 
at least in some epochs. 
Instead, they have never been observed in more than 30 years of X-rays and optical/ir
observations of blazars,  requiring a `` cosmic conspiracy" to always hide the hard emission below 
a more normal component or in seldom-observed bands (mm ?). 
Intriguingly, there was a recent claim of possible observational evidence for a low-energy cutoff 
in the Swift data of 1ES 0229+200, as an unusually high X-ray-to-UV flux ratio (\cite{tav0229}). 
However, a bug was found in the calculation of the effective area of the 
X-ray data, and the correct result (see Fig. \ref{fig3}) show a spectrum in line with 
all other objects and observations.  Future observations with ALMA (mm) and NuSTAR (hard X-ray) 
may provide further insights.  Dropping the requirement  of a hard particle spectrum, 
internal absorption on a narrow-banded photon field can make the gamma-ray spectrum 
extremely hard (\cite{intabs})
However, large fluxes in the GeV range are required,  and these seem now excluded by the
low fluxes or non-detection of the hard-TeV sources with Fermi-LAT (\cite{unolat}).

The second reason for a low EBL density (and thus for $\Gamma_{int}\gtrsim 1.5$) 
is that a high EBL would require a dramatic change of blazar properties in a very narrow
range of redshifts (see Fig. \ref{fig3}).  A high EBL would create a sharp dicothomy
in blazars spectra around $z=0.15$: all sources below this redshift  
would have $\Gamma\geq1.5$, while all blazars above this redshift would have always $\Gamma<1.5$, 
all other properties being equal or very similar.  There is no known reason, observational bias
or evolution parameter  that can explain such an abrupt change of properties in such small redshift range.  
Instead, a low EBL intensity makes the range of TeV blazar spectra 
consistent among each other  at any redshift sampled so far.

Therefore, though not yet ``bullet-proof",
until proven otherwise a low EBL seems the preferable solution according to Occam's razor.
(see also \cite{madausilk} for the cosmological problems to explain a high EBL flux).

\begin{figure*}[t]
\begin{center}
 \includegraphics[width=0.45\textwidth]{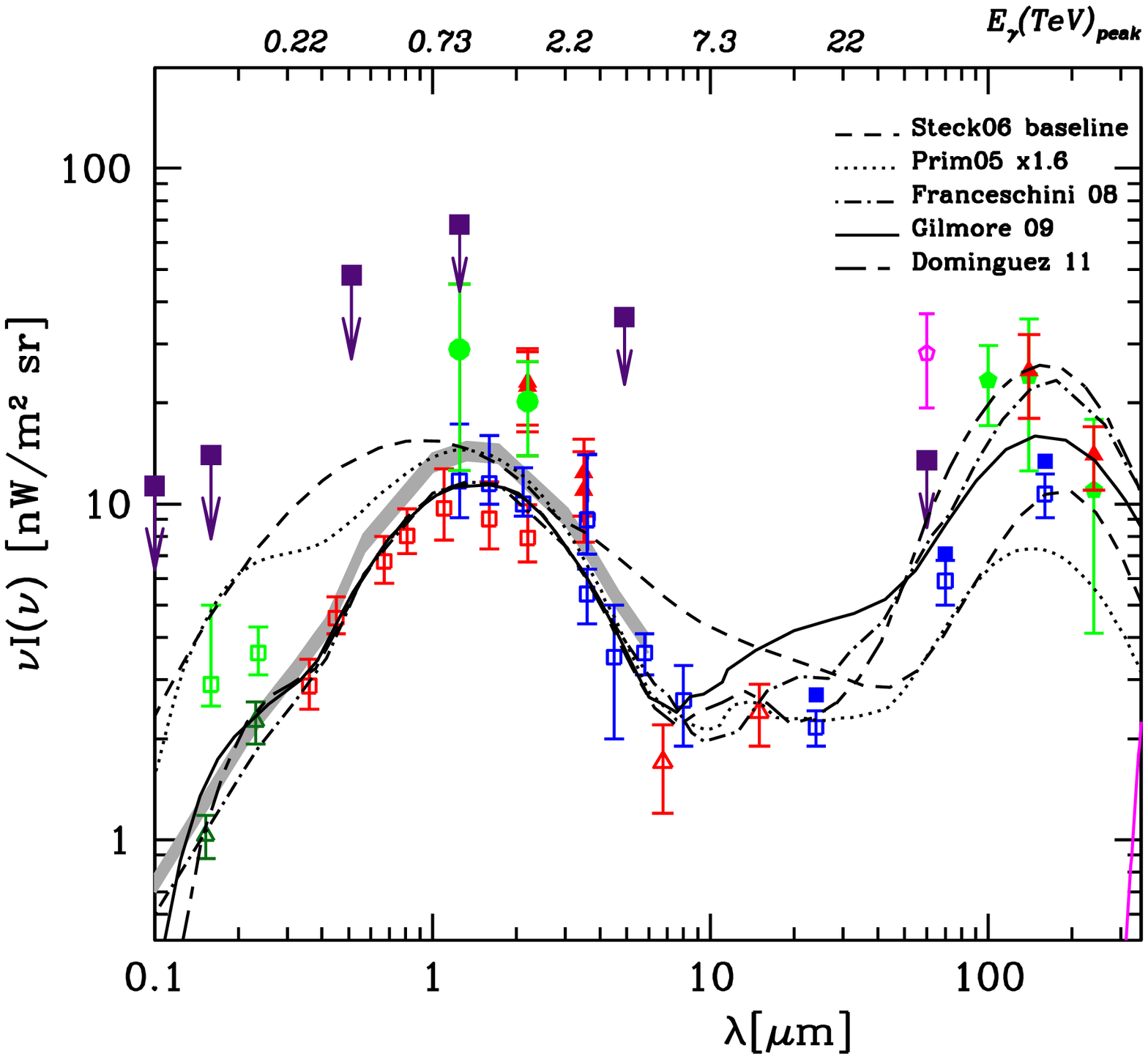}  
 \includegraphics[width=0.43\textwidth]{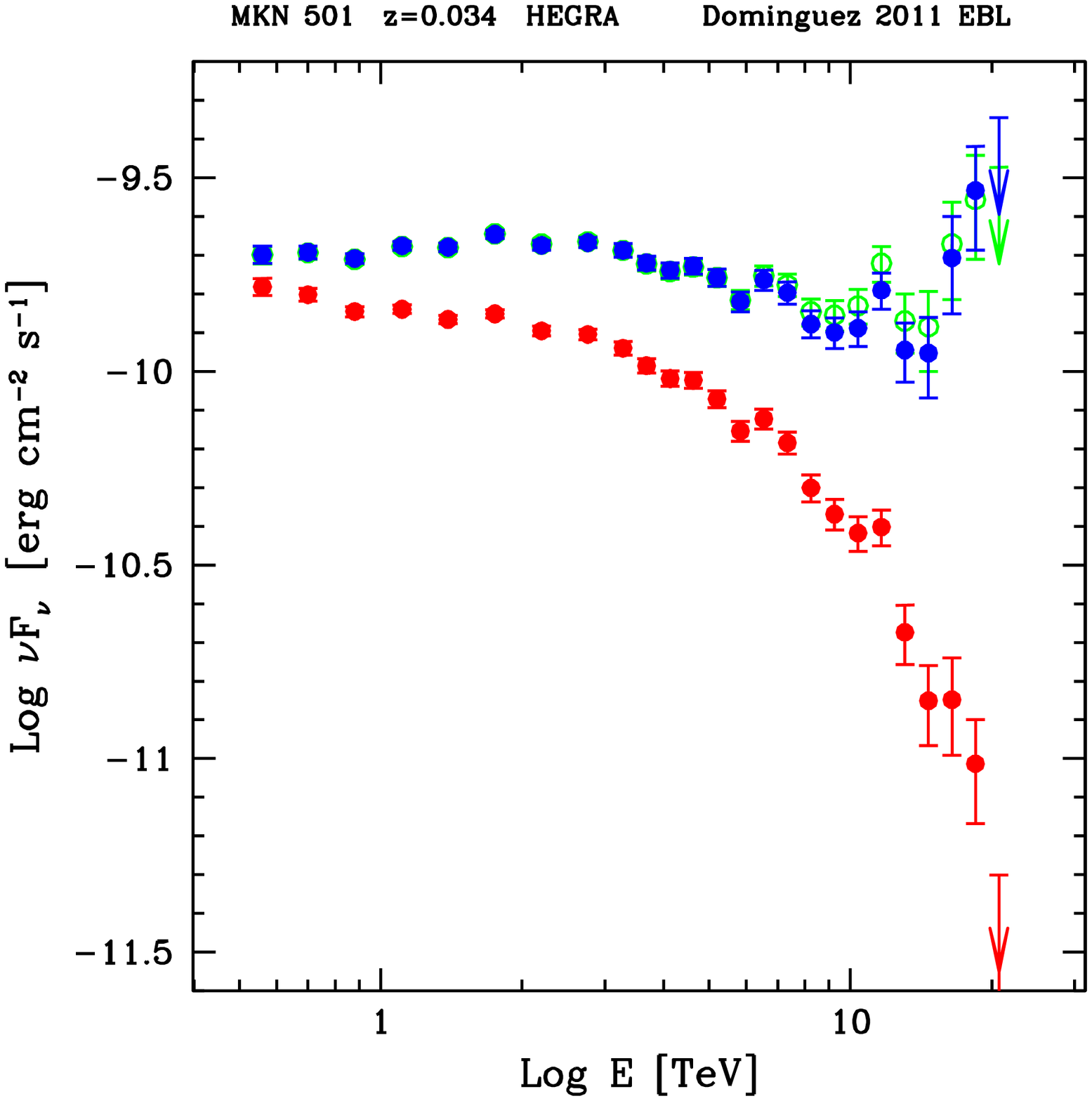} 
 \vspace*{-0.2 cm}
 \caption{Left: SED of the EBL with the most recent models. The grey band
show the EBL upper limits in the Opt-NIR range
(with the $+1\sigma$ statistical uncertainty from the VHE data)
re-calculated using the EBL shape by \cite[Franceschini et al. (2008)]{franceschini}.
Right: HEGRA spectrum of Mkn 501, as observed (red, lower points) and absorption-corrected
with two recent EBL-model calculations (\cite{dominguez} and \cite{gilmore11}, upper black-blue and grey-green 
points, respectively). Both cases show an upturn.}
   \label{fig4}
\end{center}
\end{figure*}

\section{The Chain of Constraints}
Since the EBL photon field is the same for all sources, and evolving in redshift,
the spectral-hardness limits from different blazars and in different energy bands 
can be combined to constrain the EBL over a wide range of wavelengths, and in a
more stringent way than allowed by each single object
(\cite{costamante1,dwek05,mazinraue}).
Each new spectrum can take advantage of the previous limits, forming
a chain of constraints that starts with (and depends on) the assumption on the UV flux.
With the reasonable hypothesis that the UV background  is lower than the Opt-NIR one
(e.g. around 2-6 nW/m2 sr),  blazars spectra between 0.1 and 1 TeV pin down the
Opt-NIR flux very close to the galaxy counts limit (\cite{nature}).

With this limit, the  blazar spectra between $\sim1$ and 10 TeV 
(1ES 0229+200 \cite{0229} and 1ES 1426+428, \cite{hegra1426}) constrain the EBL spectrum  
between 2 and 10 \m\  to a slope close to $\lambda^{-1}$, again very close to the 
lower limits from galaxy counts  (\cite{0229}).
This is possible because the Opt-NIR limit precludes the possibility 
of increasing the flux at $\sim$1 \m\ 
to reduce the difference in optical depth between 1 and 10 TeV (see Sect. 2).

The consequent upper limit at 10 \m\  constrains the rising of the EBL SED
towards the far-infrared hump, which is determined by warm and cold dust emission.
This band is probed by $\gamma$-rays  between $\sim8$ and 100 TeV.
So far, only two close-by objects have been measured up to 15-20 TeV, 
thanks to HEGRA observations:  Mkn 421 ($z$=0.031, \cite{hegra421}) and Mkn 501 
($z$=0.034, \cite{hegra501}).  
Interestingly, the HEGRA spectrum of Mkn 501 seems to have problems with the most 
recent EBL calculations, which cause an up-turn or pile-up at the highest energies
(see Fig. \ref{fig4}). This is the modern version of the so called ``TeV-FIR background crisis"
(see e.g. \cite{felix_icrc}), caused years ago by a first very high estimate of the 50 \m\ EBL flux.
However the information on warm dust and the $\gamma$-ray statistic
are still insufficient to draw any sensible conclusion. 
More data $>10$ TeV are needed (hopefully from CTA).

\section{Future and Conclusions}
VHE data sample mainly the local EBL ($z=0-0.5$).
To probe the EBL evolution over cosmic time and in the UV band,
sources up to $z=3-4$ and data in the 10-100 GeV range are needed. 
The first results from Fermi-LAT (\cite{fermiebl}) are in agreement with the VHE limits and 
with the most recent EBL evolution models.  To this respect, however, GRBs might turn out to be 
more useful $\gamma$-ray sources, given that EBL and blazar-intrinsic evolutionary effects
might be very difficult to disentangle (\cite{anita}).

At VHE, instead, CTA is NOT expected to improve the Opt-NIR limits per se, contrary to various claims.
The reason is twofold. First, there is not much room for improvement ! 
The present upper limits  already match the lower limits by galaxy counts quite closely (see Fig. \ref{fig4})
Second, even with infinite $\gamma$-ray statistics, it remains the unavoidable systematic uncertainty 
in blazar modeling. The small $\Delta\Gamma\sim0.1-0.3$ induced by the residual EBL uncertainty
between lower and upper limits can be typically accomodated with very small changes in blazar parameters,
for a given set of multiwavelength data.

The main question to be addressed is if our blazar assumptions are correct 
(and in such case the EBL is already pinned down) or are not (and in this case by how much). 
In the latter case, the entire construction must be revisited,
since it is our understanding of the gamma-ray sources that changes completely.
It is on this aspect, on blazar physics, that CTA is expected to provide the most significant improvements
and to test our assumptions by finding counter-examples. This can be done by monitoring more 
low-redshift objects in the TeV range (to detect directly $\Gamma<1$) and by measuring more
high-redshift spectra (to find those for which $\Gamma=1.5$ is incompatible
with galaxy counts; \cite{barca}).

In conclusion, a low EBL close to galaxy counts seems the ``convergent solution",
despite some uncertainties on blazar physics.  However, there are still some fundamental aspects
of blazars' acceleration and emission mechanisms not yet understood, 
which present and future observations are testing.
The EBL limits obtained so far are robust against small changes in blazars assumptions,
but they all depend, by construction,  on the assumption of a low diffuse UV flux. \vspace{0.9cm}

{\bf \noindent ACKNOWLEDGMENTS}: 

I would like to thank the Conference Organizers for the 
invitation and financial support, and the Max-Planck-Institut f\"ur Kernphysik
for the ospitality and support during my visit.

\bigskip
\noindent {\bf DISCUSSION}

\bigskip
\noindent {\bf Q (K. Mattila):} 
Does MAGIC give additional (or better) constraints to the EBL at optical wavelengths, 
where it should be more sensitive than HESS because of its lower $\gamma$-ray energy range ?
(I am referring to the 2008 Science paper on 3C279).

\bigskip
\noindent {\bf A:} 
In this case (3C 279), not really. The problem is the much higher uncertainty in the spectrum.  
This is also caused by 1) the smaller detected band (less than half a decade, with respect to a full 
decade in energy for 1ES\,1101-232 and the other BL Lacs), and
2) the much lower S/N.  The lower $\gamma$-ray energy range does probe slightly shorter wavelengths,
but it also reduces the sensitivity of the $\gamma$-ray spectrum to EBL changes, 
with respect to the 0.2-2 TeV range, despite the higher redshift. 
Note that in the blue-UV range ("max EBL" curve in their Fig. S2), 
the given values are basically arbitrary, and should not be considered limits at all 
(a higher UV flux makes also the Magic spectrum softer,  see Sect 2).
In summary, that observation  does not provide more stringent constraints on the optical background, 
though the detection itself corroborates the idea of a low overall EBL 
(adopting historical GeV peak fluxes for 3C\,279).
It is also important to remind that, for this source, there is the further systematic uncertainty 
of internal absorption on UV photons from the broad line region 
(this object has broad lines in the optical spectrum).

\end{document}